\documentclass[pre,aps,10pt,superscriptaddress,twocolumn,floatfix]{revtex4-2}

\usepackage[nice]{nicefrac}
\usepackage{graphicx,color}
\usepackage{amsmath,amssymb,bm,bbm}
\usepackage{amsmath}
\usepackage{braket}
\usepackage{bbold}
\usepackage{float}
\usepackage{physics}

\newcommand{\ie}{{i.e.},\ }

\usepackage[plainpages=false,pdfpagelabels,colorlinks=true,linkcolor=red,urlcolor=blue,citecolor=blue,pdftitle={},pdfauthor={},pdfdisplaydoctitle=true,pdfduplex=DuplexFlipLongEdge]{hyperref}

\definecolor{darkred}{rgb}{0.90,0.2,0.2}
\definecolor{darkgreen}{rgb}{0,0.60,.2}
\definecolor{darkblue}{rgb}{0.1,0.3,1}
\definecolor{grey}{cmyk}{0,0,0,0.25}
\definecolor{orange}{cmyk}{0,0.6,0.8,0}

\usepackage[T1]{fontenc}

\begin{document}
\title{Scaling Theory of Fading Ergodicity}

\author{Rafał Świętek}
\affiliation{Department of Theoretical Physics, J. Stefan Institute, SI-1000 Ljubljana, Slovenia}
\affiliation{Department of Physics, Faculty of Mathematics and Physics, University of Ljubljana, SI-1000 Ljubljana, Slovenia\looseness=-1}

\author{Miroslav Hopjan}
\affiliation{Department of Theoretical Physics, J. Stefan Institute, SI-1000 Ljubljana, Slovenia}
\affiliation{Institute of Theoretical Physics, Wroclaw University of Science and Technology, 50-370 Wrocław, Poland}

\author{Carlo Vanoni}
\affiliation{SISSA – International School for Advanced Studies, via Bonomea 265, 34136, Trieste, Italy}
\affiliation{Department of Physics, Princeton University, Princeton, New Jersey, 08544, USA}

\author{Antonello Scardicchio}
\affiliation{INFN Sezione di Trieste, Via Valerio 2, 34127 Trieste, Italy}
\affiliation{ICTP, Strada Costiera 11, 34151, Trieste, Italy}

\author{Lev Vidmar}
\affiliation{Department of Theoretical Physics, J. Stefan Institute, SI-1000 Ljubljana, Slovenia}
\affiliation{Department of Physics, Faculty of Mathematics and Physics, University of Ljubljana, SI-1000 Ljubljana, Slovenia\looseness=-1}

\begin{abstract}
In most noninteracting quantum systems, the scaling theory of localization predicts one-parameter scaling flow in both ergodic and localized regimes.
A corresponding scaling theory of many-body ergodicity breaking is still missing. Here, we introduce a scaling theory of ergodicity breaking in interacting systems, in which the divergent relaxation time follows from the Fermi golden rule, and the observable fluctuations in proximity of the ergodicity breaking critical point are described by the recently introduced fading ergodicity scenario.
We argue that, in general, the one-parameter scaling is insufficient, and we show that the scaling theory predicts the critical exponent $\nu = 1$ at the ergodicity breaking critical point.
Our theoretical framework may serve as a building block for two-parameter scaling theories of many-body systems.
\end{abstract}

\maketitle

\textit{Introduction---}
The emergence of thermalization in isolated quantum many-body systems has attracted considerable attention since the beginning of quantum mechanics.
While it is nowadays understood that in most isolated quantum many-body systems, local observables approach the thermal ensemble predictions after being taken out of equilibrium~\cite{deutsch_91, srednicki_94, srednicki_99, rigol_dunjko_08, dAlessio16, mori_ikeda_18, deutsch_18, Trotzky2012, langen_schmiedmayer_13, Eisert2015, mbl_review}, much less is known about typical (as opposed to integrable) systems that avoid thermalization in the thermodynamic limit.
If, upon tuning a parameter, the system can be made to switch from a thermal to a nonthermal phase, of particular importance becomes the understanding of the critical behavior at the boundary, and its relationship to other critical phenomena such as ground-state quantum phase transitions~\cite{sachdevbook} or spin-glass transitions \cite{binder1986spin, mezard1987spin}.

Since most of the studies are carried out in finite systems, the central question concerns how the properties of those systems flow with increasing the system size either towards ergodicity, for which the eigenstate thermalization hypothesis (ETH) represents a fixed point~\cite{deutsch_91, srednicki_94, srednicki_99, rigol_dunjko_08, dAlessio16}, or towards localization, which is another fixed point.
A paradigmatic example of the latter is Anderson localization~\cite{Anderson65}, which became a hallmark of ergodicity breaking in noninteracting quantum systems.
A milestone in the understanding of Anderson transition in finite-dimensional lattices has been the scaling theory by Abrahams, Anderson, Licciardello, and Ramakrishnan~\cite{anderson79}, which is based on the one-parameter scaling hypothesis, investigated numerically by later studies~\cite{MacKinnon81, MacKinnon83}. Recent studies have further developed this theoretical framework using modern observables, {\it e.g.}, spectral statistics, wave function fractal dimensions, and entanglement entropy~\cite{slevin_ohtsuki_99,Rodriguez10, Lopez_delande_12, Slevin_2014, tarzia_tarquini17, devakul_huse_2017, slevin_ohtsuki_18, luo_ohtsuki_2022, Sierant2020, suntajs_prosen_21, suntajs_prosen_23, vanoni_scardicchio_2024, altshuler_vanoni_2024}, and they explored its validity and limitations in special geometries, such as random regular graphs~\cite{garciamatia_lemari_17,garciamatia_lemari_20,pino_20,garciamata2022,sierant_scardicchio_23,vanoni_scardicchio_2024}. In such graphs, it was found \cite{altshuler_vanoni_2024} (see also \cite{garciamata2022,sierant_scardicchio_23}) that the one-parameter scaling hypothesis is not sufficient, and one needs to resort to a two-parameter scaling theory, similar in many ways to what describes the Kosterlitz-Thouless transition \cite{kosterlitz2016kosterlitz}.

A common perspective on many-body quantum systems, which are candidates for ergodicity breaking, is that they exhibit many analogies with the Anderson model on expander graphs~\cite{altshuler1997quasiparticle, Basko06, sierant_scardicchio_23, vanoni_scardicchio_2024, roy_logan_24}.
Yet, they do not seem to be governed by the one-parameter scaling hypothesis.
The evidence supporting this expectation is diverse~\cite{suntajs_bonca_20a, suntajs_bonca_20b, leblond_sels_21, kim_polkovnikov_24, bhattacharjee_adreanov_24}, and in most cases, it is obtained by studying the random-field XXZ model~\cite{suntajs_bonca_20a, suntajs_bonca_20b, Panda20, sierant_lewenstein_20, Laflorencie20, sels2020, sels_polkovnikov_2023}; see~\cite{niedda2024renormalization} for a recent example.
In the latter model, the many-body ergodicity-breaking transition is the many-body localization transition~\cite{oganesyan_huse_07, Znidaric08, de2013ergodicity,  Alet18, abanin2019, mbl_review}.

However, instead of following a specific model such as the random-field XXZ model, we here consider a more general theory of many-body ergodicity breaking, whose main features are described by rather simple theoretical arguments. In particular, we consider systems in which the divergent relaxation time close to the ergodicity breaking critical point obeys the Fermi golden rule~\cite{luitz_huveneers_17, suntajs_vidmar_22}.
When considering physical observables, this framework naturally gives rise to a scenario of many-body ergodicity breaking dubbed fading ergodicity~\cite{kliczkowski_vidmar2024}. The latter is an ergodic regime in which the ETH ceases to be valid in its conventional form~\cite{srednicki_99, rigol_dunjko_08, dalessio_kafri_16}.
A toy model of fading ergodicity is the quantum sun model~\cite{suntajs_vidmar_22, suntajs_deroeck_24}, which is closely connected to the avalanche theory in spin-1/2 chains~\cite{deroeck_huveneers_17, luitz_huveneers_17, thiery_huveneers_18, altman_potirniche_19, crowley_chandran_20, suntajs_vidmar_22, crowley_chandran_22b, Sels_2022, Morningstar2022, leonard_greiner_2023, suntajs_deroeck_24, pawlik_zakrzewski_2024, szoldra_sierant_24}. 

\begin{figure}[t!]
\centering
\includegraphics[width=\columnwidth]{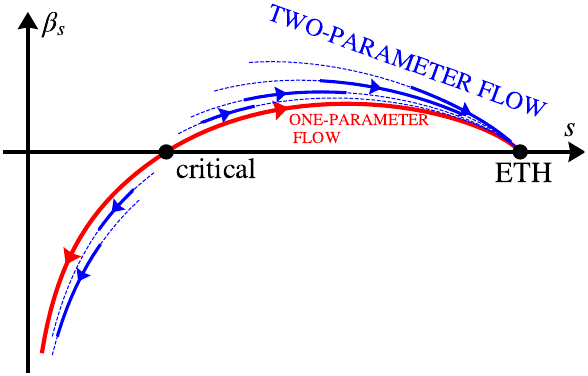}
\caption{
Two-parameter scaling of fading ergodicity.
Dashed blue lines are results for the beta function $\beta_s$ from Eq.~\eqref{eq:beta:complete} at different interactions $\alpha$, and the overlapping solid blue lines are results for the numerically available system sizes $L$.
The solid red line is the critical one-parameter flow with the critical exponent $\nu=1$, \ie Eq.~\eqref{eq:beta:complete} with $a=a_c$.
The two-parameter flows in the ergodic phase terminate, with the same derivative, at the ETH fixed point $s=1$.
}
\label{figM1}
\end{figure}

In this Letter, we introduce a scaling theory of fading ergodicity and argue that it generally gives rise to a two-parameter scaling. Studying the entanglement properties of many-body quantum states, we inspect the flow in the vicinity of the two special points, i.e., the ergodicity breaking critical point~\cite{suntajs_vidmar_22, pawlik_zakrzewski_2024, suntajs_deroeck_24} and the ETH fixed point~\cite{kliczkowski_vidmar2024}, as sketched in Fig.~\ref{figM1}. 
We show that the two-parameter description contains a rather simple structure, and it predicts the critical exponent $\nu = 1$ at the ergodicity breaking critical point. Our results suggest a different scenario from that depicted by the random-field XXZ model, and present a new universality class of ergodicity-breaking transitions.

\textit{Setup---}
Our goal is to develop a scaling theory of ergodicity breaking transition in spin-1/2 systems.
While the theory is based on general arguments, a particular system we have in mind for numerical tests is the quantum sun model~\cite{suntajs_vidmar_22, suntajs_deroeck_24, luitz_huveneers_17, crowley_chandran_20}.
Its Hamiltonian is given by $\hat H = \hat R + g_0 \hat H_{\rm int} + \hat H_{\rm loc}$, in which $\hat R$ describes a quantum dot of dimension $2^N$ (with fixed $N=3$ in our case), $g_0 \hat H_{\rm int}$ represents the interaction between the $N$ spin-1/2 qubits within the dot and $L$ spin-1/2 qubits outside the dot, and $\hat H_{\rm loc}$ describes the random fields of $O(1)$ strength.
Full details of $\hat H$ are given in Eq.~(\ref{eq:qsun}) of End Matter.
Two parameters are needed to describe this Hamiltonian: the overall interaction strength $g_0$, and the interaction $\alpha\in[0,1]$ that determines the decay factor for each spin coupling, $\hat H_{\rm int} \approx \sum_{\ell=1}^L \alpha^\ell \hat S_{n(\ell)}^x \hat S_\ell^x$, where $\ell$ labels the spin-1/2 outside the dot and $n(\ell)$ labels, for a given $\ell$, a randomly chosen spin-1/2 inside the dot.
The coupling term $\hat H_{\rm int}$ can be interpreted as the exponentially decreasing interaction between two spins at distances $\ell=1,...,L$. We allow for additional fluctuations of the distance around the integers, $\ell \rightarrow u_\ell$, with $u_\ell\in[\ell-0.2,\ell+0.2]$, which account for small uncertainties in the particle positions but do not affect the properties of the ergodicity breaking transition.
A critical $\alpha_c$ separates an ergodic phase at $\alpha>\alpha_c$ from a localized phase at $\alpha<\alpha_c$, with $\alpha_c \approx 1/\sqrt{2}$. 
Fading ergodicity is observed at $\alpha_c < \alpha \lesssim 1$, while the observable fluctuations are close to those predicted by the conventional ETH at $\alpha \approx 1$~\cite{kliczkowski_vidmar2024}.

We identify the single-site entanglement entropy as an observable with well-defined properties in the ergodic and localized regimes.
We consider the bipartite entanglement entropy in an eigenstate $\ket{n}$, partitioned into a subsystem $A$, being a single spin at distance $\ell\sim L$ from the dot, and its complement $B$. We define the reduced density matrix as $\hat{\rho}_A^n=\Tr_B\dyad{n}$ and compute the von Neumann entanglement entropy $S^n_A=-\Tr\qty(\hat{\rho}_A^n\ln{\hat{\rho}_A^n})$, which we average over eigenstates in a microcanonical energy window at the mean total energy, denoted as $\overline{S}_A$. 
We define the scaled average entanglement entropy as $s = \overline{S}_A/S_A^{\mathrm{max}}$, where $S_A^{\mathrm{max}}=\ln 2$ is the maximal entropy. 
This quantity interpolates between the localized phase, $s\to0$, and the ergodic phase, $s\to1$~\cite{suntajs_deroeck_24}.

{\it Beta function---}
The goal of the Letter is to study the scaling properties of $s$ in terms of the beta function,
\begin{equation} \label{def_beta_s}
\beta_s = \frac{d\ln{s}}{d\ln{L}} \;.
\end{equation}
The beta function was originally introduced for describing the scaling behavior of the dimensionless conductance $g$~\cite{anderson79}. Under the one-parameter scaling hypothesis, which is expected to hold in the $d$-dimensional Anderson model, it can be used to obtain the scaling for any observable $t$, by means of universal change of variables, $t(g)$, see \cite{altshuler_vanoni_2024}. We consider here, instead, the beta function of the wave function entanglement entropy. For the latter (and also for other indicators such as the wave function fractal dimension or spectral statistics~\cite{vanoni_scardicchio_2024, altshuler_vanoni_2024, kutlin2024investigating}), the beta function $\beta_s$ displays three fixed points: the ergodic (ETH) fixed point at $s=1$, the localized fixed point at $s=0$, and a third isolated zero at the critical point $s_c$ (Wilson-Fisher fixed point), with $0<s_c<1$.

When a critical point $\alpha=\alpha_c$ is described by a one-parameter scaling, the beta function is expected to display power-law corrections $L^{-\gamma}$ to the critical entanglement entropy $s=s_c$.
In the vicinity of the critical point, one can linearize the beta function,
\begin{equation}\label{eq:beta:def}
    \beta_s=b_0 (s-s_c) + O((s-s_c)^2) \;,
\end{equation}
such that $b_0 = \beta_s'$ at $s=s_c$.
Solving Eq.~\eqref{eq:beta:def} for $s$ produces the scaling relation
\begin{equation}\label{eq:sp:ansatz}
    s(L)=s_c+f((L/\xi)^{1/\nu})\;,
\end{equation}
where $\nu$ is the critical exponent, $\nu=1/(s_c b_0)$, and $f(z)$ is a scaling function that is linear for small $z$~\cite{niedda2024renormalization}. 
The correlation or localization length $\xi$ diverges at the critical point.
Previous works studied numerical scaling collapses using Eq.~(\ref{eq:sp:ansatz}) for the level spacing ratio~\cite{suntajs_vidmar_22, pawlik_zakrzewski_2024}, and in End Matter we show similar scaling collapses for $s$.

{\it Scaling arguments---}
Let us first focus on the behavior of the beta function~\eqref{def_beta_s} in the ergodic regime.
To do so, we need to describe the dependence of $s$ on the linear size $L$. If the entire ergodic phase is described by the conventional ETH, the approach to the ETH fixed point $s=1$ is governed by random-matrix theory corrections of the form $s\simeq 1 - c/D+O(D^{-2})$, where $D\propto 2^L$ is the Hilbert-space dimension.
Such corrections are typical for Haar-random pure states~\cite{bianchi_hackl_22}, see also~\cite{SM}.
However, in the fading ergodicity regime when the conventional ETH breaks down~\cite{kliczkowski_vidmar2024}, one expects the leading correction to the entanglement entropy $s$ to scale as
\begin{equation}\label{eq:s:corrections}
    s = 1-{c}/{D_{\rm eff}}=1-ce^{-L/\eta}\,,
\end{equation}
with the fluctuation exponent $\eta$ depending on $\alpha$, approaching $\eta = 1/\ln 2$ at the ETH point ($\alpha\to 1$) and $\eta\to\infty$ at the ergodicity breaking critical point ($\alpha\to\alpha_c$). 
The ansatz in Eq.~\eqref{eq:s:corrections} is motivated by the similarity of the scaling properties of the single-site entanglement entropies and the variances of the matrix elements of local observables introduced in~\cite{kliczkowski_vidmar2024}, see also~\cite{SM} for details.
Indeed, it was argued~\cite{kliczkowski_vidmar2024} that the fluctuation exponent $\eta$ scales as $\eta^{-1}=\ln{2}(1-\ln\alpha/\ln\alpha_c)$, up to a constant prefactor.
Hence, it diverges at the critical point and it assumes the role of $\xi$ from Eq.~(\ref{eq:sp:ansatz}).

The scaling ansatz from Eq.~(\ref{eq:s:corrections}) is the central building block of our theory.
It corresponds to the following beta function,
\begin{eqnarray}\label{eq:beta:ergodic}
    \beta_{\rm fading}(s,c)&=& -\frac{1-s}{s}\ln(1-s)+\frac{1-s}{s}\ln(c) \;.
\end{eqnarray}
The main result of Eq.~\eqref{eq:beta:ergodic} is that, if $c$ depends on a microscopic parameter, such as $\alpha$, then $\beta_{\rm fading}$ is a two-parameter function.
In the quantum sun model, we show that, indeed, $c$ depends on the interaction $\alpha$.

For a complete description of the renormalization group flow close to ergodicity breaking, we need to incorporate also the behavior in the localized phase. Similarly to the fractal dimension in the Anderson model and to other spectral observables \cite{vanoni_scardicchio_2024}, the entanglement entropy decreases exponentially as $s = b e^{-L/\eta_{\rm loc}}$ in the localized phase. This behavior corresponds to a beta function
\begin{equation}\label{eq:beta:loc}
    \beta_{\rm loc}(s,b) = \ln{s}-\ln{b}\,,
\end{equation}%
which is consistent with the localized fixed point $\beta(s\to0)\to-\infty$.
If the constant $b$ depends on interaction $\alpha$ (we argue it does), it leads to a two-parameter flow. 

Following similar arguments as presented in~\cite{anderson79}, we argue that the beta function in Eqs.~\eqref{eq:beta:ergodic} and~\eqref{eq:beta:loc} indeed ensures the ergodicity-breaking transition at finite $s_c$. On the ergodic side, the beta function approaches the ETH fixed point with a negative derivative (\ie from above), while close to the localized fixed point it is negative. Consequently, there has to occur a sign change in the beta function at $s_c=s(\alpha_c)$, where $0<\alpha_c<1$, determining the transition from ergodicity to localization, see Fig~\ref{figM1}.%

\begin{figure}[t!]
\centering
\includegraphics[width=\columnwidth]{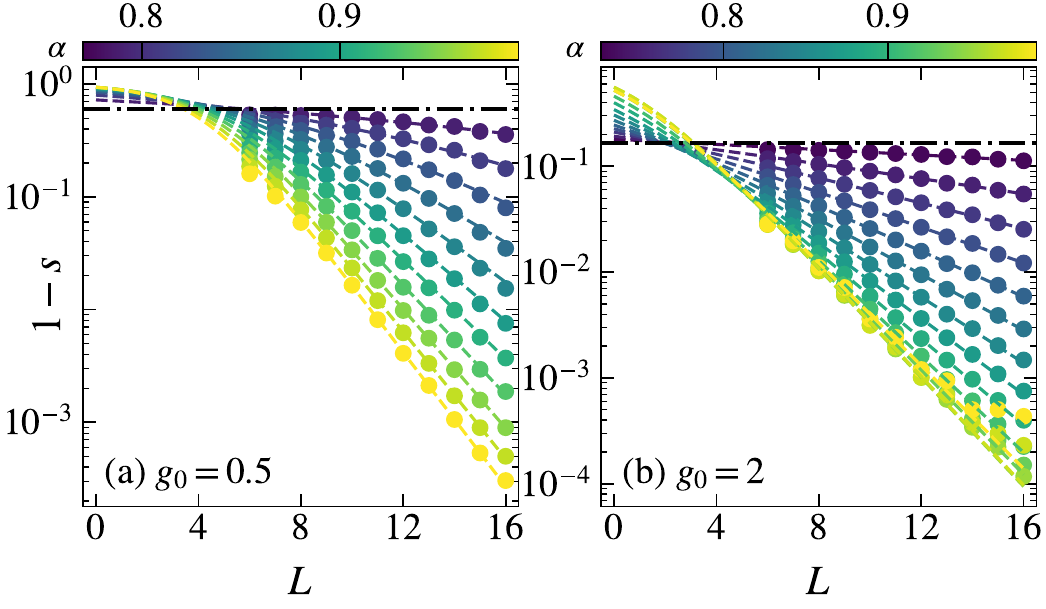}
\caption{
Symbols: Numerical results for $1-s$ in the ergodic phase of the quantum sun model at {  (a) $g_0=0.5$ and (c) $g_0=2$}.
Dashed lines are fits using Eq.~(\ref{eq:s_Fermi}), for system sizes $L>9$, with free parameters $a$ and $\eta$,  see~\cite{SM} for the actual values of $a$ and $\eta$.
The horizontal dash-dotted lines are the values $1-s_c$ at the critical point.
}
\label{figM2}
\end{figure}

To simplify the discussion, we argue that the specific scaling behaviors on the ergodic and localized sides discussed so far can be interpreted as limiting cases to a general function.
We propose that the entanglement entropy can be modeled for any value of interaction as
\begin{equation}
\label{eq:s_Fermi}
    s(L) = \frac{1}{1+a \, e^{-L/\eta}},
\end{equation}
which incorporates both the ergodic, $\eta>0$, and localized, $\eta=-\eta_{\rm loc}<0$, limiting cases. The parameter $a$ in Eq.~\eqref{eq:s_Fermi} is related to the parameters $c$ and $b$ as $a \approx c$ on the ergodic side at $L/\eta\gg 1$, and $a \approx b$ on the localized side at $L/\eta \ll -1$.
For the quantum sun model, we demonstrate relevance of the ansatz~\eqref{eq:s_Fermi} in Fig.~\ref{figM2} {  (at $g_0=0.5,2$)}, and in Fig.~\ref{figM5} of End Matter {  (at $g_0=1$)}.
Following Eq.~\eqref{eq:s_Fermi} one is able to find the corresponding beta function, 
\begin{equation}
\label{eq:beta:complete}
    \beta_s(s, a) = -(1-s) \ln\frac{1-s}{s} + (1-s)\ln{a}\,.
\end{equation}
Remarkably, $\beta_s$ in Eq.~\eqref{eq:beta:complete} matches Eq.~\eqref{eq:beta:ergodic} for $1-s\ll 1$, and Eq.~\eqref{eq:beta:loc} when $s\ll 1$. It is a good, simple, interpolating function for $\beta$ both around the localized and ETH regimes and, in the remainder of the Letter we will prove that it also describes the vicinity of the critical point $s_c$.

The first main result of the Letter is that $\beta_s$ from Eq.~\eqref{eq:beta:complete} is a two-parameter function since the parameter $a$ depends on the interaction $\alpha$.
The dependence of $a$ on $\alpha$ is seen in the fits in Fig.~\ref{figM2} at $L\to0$, and explicitly demonstrated in Sec.~S3~B in~\cite{SM}.

We note that it appears tempting to express the beta functions in Eqs.~\eqref{eq:beta:ergodic},~\eqref{eq:beta:loc} and~\eqref{eq:beta:complete} as $\beta(s,p)=\beta_1(s)+\beta_2(s,p)$, where $p\in\{a, b, c,...\}$ is a microscopic parameter. They have in common that the two-parameter scaling part of the flow is subleading in two limits: close to the localized fixed point $s=0$, and close to the ETH fixed point $s=1$, as $\beta_2(s,p)/\beta_1(s)\to 0$ when $s\to 0$ or $s\to 1$. This means that, asymptotically close to these points, the flow returns to one-parameter scaling, which {  will} be discussed in more detail below. In this sense, our result is fundamentally different from the Kosterlitz-Thouless flow~\cite{kosterlitz2016kosterlitz} or that of the Anderson transition on random graphs~\cite{vanoni_scardicchio_2024}.
Nevertheless, it is crucial to consider the complete beta function for understanding the existence and properties of the critical point separating the localized and ETH fixed points, since neglecting the function $\beta_2(s,p)$ yields a different critical exponent $\nu$.

The second main result of the Letter concerns the value of the critical exponent $\nu$ in Eq.~\eqref{eq:sp:ansatz}.
At the critical point $s_c$, the entanglement entropy is $s_c=1/(1+a_c)$, where $a_c\equiv a(\alpha_c)$, as it is easily seen either by solving $\beta_s(s,a)=0$ or by letting $\eta\to\infty$ in Eq.~(\ref{eq:s_Fermi}).
The slope of the beta function at $s=s_c$ is $\beta_s'(s=s_c,a_c)=1/s_c$, and $\nu$ is
\begin{equation} \label{def_nu_1}
    \nu=\qty(\beta_s'\, s_c)^{-1}=1\;.
\end{equation}
This family of $\beta_s(s,a)$ all have $\nu=1$, irrespective of $g_0$, \ie of the value of $s_c$. Hence, $\beta_s(s,a)$ describe a family of flows in the same universality class, irrespective of $a$.

The result $\nu=1$ is consistent with previous studies~\cite{deroeck_huveneers_17, suntajs_vidmar_22, crowley_chandran_22b}, e.g., with the divergence of the ratio of the Heisenberg and Thouless times close to the critical point~\cite{suntajs_vidmar_22}.
In fact, the ratio was shown to diverge as $\propto e^{L/\eta}$, with $\eta$ introduced above.

\begin{figure}[t!]
\centering
\includegraphics[width=\columnwidth]{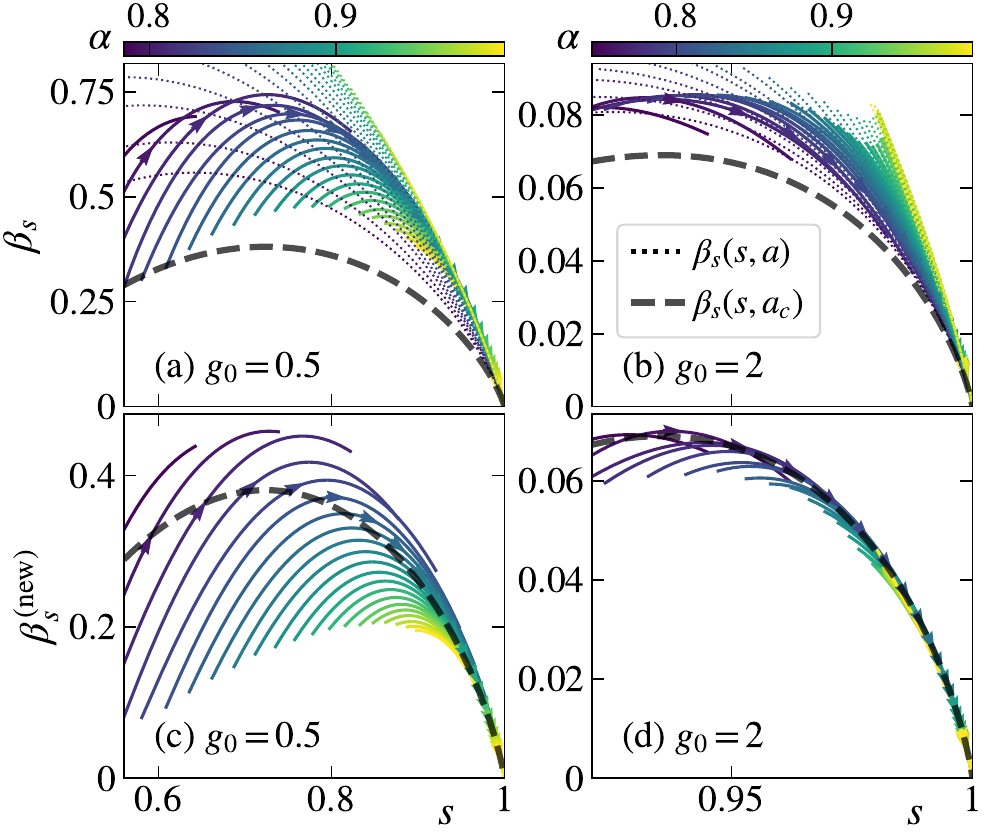}
\caption{ 
{  (a),(b)} Results for $\beta_s$ at (a) $g_0=0.5$ and (b) $g_0=2$.
{  (c),(d)} Results for $\beta^{\rm (new)}_s$ at (c) $g_0=0.5$ and (d) $g_0=2$.
All results correspond to the ergodic phase $\alpha>\alpha_c$ in the quantum sun model.
Solid lines are numerical results for $\beta_s$ in {  (a),(b)}, see Eq.~\eqref{def_beta_s}, and for $\beta^{\rm (new)}_s$ in {  (c),(d)}, see Eq.~\eqref{eq:beta:new}, in which the value of $L_0$ is estimated from the fits in Fig.~\ref{figM2} (see also~\cite{SM} for details of numerical calculations).
The arrows indicate the increase in system size $L=7,...,16$, and the colors denote the value of interaction $\alpha$. 
Thin dotted lines in {  (a),(b)} show the two-parameter functions $\beta_s(s,a)$ from Eq.~\eqref{eq:beta:complete}, with values of $a$ extracted from Fig.~\ref{figM2}, while the thick dashed (black) lines show the one-parameter function $\beta_s(s,a=a_c)$. 
The corresponding values of $\alpha$ are the same in all panels, and the legend in panel (b) applies to all panels.
}
\label{figM3}
\end{figure}

\textit{Corrections to one-parameter scaling in the ergodic phase---}%
While the analysis of Eqs.~\eqref{eq:s:corrections}--\eqref{def_nu_1} represents the main results of the Letter, we can take a step further and define an appropriate scaling variable that considerably reduces the two-parameter flow and, more importantly, provides finite-size corrections to the critical exponent $\nu$.

To this end, we observe that the numerical results in Fig.~\ref{figM2} can also be well described by a modified Eq.~\eqref{eq:s_Fermi}, $s = (1+ a_c \exp[-(L-L_0)/\eta])^{-1}$, in which $L_0 = O(1)$ is a characteristic length that is, to a reasonable approximation, independent of $\alpha$, and the parameter $a_c$ is a constant. [The modified $s$ can be thought of as $s$ from Eq.~\eqref{eq:s_Fermi}, in which $a\to a_c e^{L_0/\eta}$.]
Then, introducing a variable $L' = L-L_0$, one defines the new beta function $\beta_s^{(\rm new)}$,
\begin{equation}\label{eq:beta:new}
    \beta_s^{(\rm new)} = \frac{d\ln s}{d\ln L'}=\qty(1-\frac{L_0}{L})\beta_s(s,a)\;.
\end{equation}
The important property of $\beta_s^{(\rm new)}$ is that it is a one-parameter function.
It is given by $\beta_s^{(\rm new)} = \beta_s(s,a_c)$, i.e., by $\beta_s$ in Eq.~\eqref{eq:beta:complete} at $a=a_c$.
Then, one can use Eq.~\eqref{eq:beta:new} to express the two-parameter function $\beta_s$ via the one-parameter function $\beta_s^{\rm (new)}$ and the $L_0/L$ correction,
\begin{equation} \label{def_beta_tps}
    \beta_s(s,L) = \beta_s(s,a_c)\qty(1-\frac{L_0}{L})^{-1} \!\! \approx \beta_s(s,a_c) + \frac{\beta_s(s,a_c)\, L_0}{L}\;,
\end{equation}
where in the last equation we expressed $\beta_s(s,L)$, at $L\gg L_0$, as a sum of the one- and two-parameter scaling part.

Numerical results in Fig.~\ref{figM3} show that, while $\beta_s$ is not a one-parameter function {  [Figs.~\ref{figM3}(a)--\ref{figM3}(b)]}, $\beta_s^{\rm(new)}$ is much closer to a one-parameter function {  [Figs.~\ref{figM3}(c)--\ref{figM3}(d)]}, in particular for $g_0\gtrsim 1$. 
The suppression of the two-parameter scaling obtained by this simple change of variables is most clearly visible when comparing {  Figs.~\ref{figM3}(d) and~\ref{figM3}(b)} at $g_0=2$. 
Indeed, for $g_0=2$ we observe a remarkable collapse of the data into a single curve, see {  Fig.~\ref{figM3}(d)}, except for the initial parts of the flow trajectories that are outside of the scaling regime. The latter represent finite-size effects that become more severe with decreasing $g_0$.
For $g_0=0.5$, see {  Fig.~\ref{figM3}(c)}, the collapse of the data is less clear. We interpret it as evidence of significant finite-size effects in the regime $g_0 < 1$, {  in which the hybridization between the ergodic quantum dot and the spins outside the dot is weaker than in the regime $g_0>1$.
However, it is expected that ergodic properties in the regime $g_0 < 1$ will be enhanced for larger system sizes than those considered numerically~\cite{suntajs_deroeck_24}.
We therefore also expect the results for $\beta_s^{\rm(new)}$ to approach the one-parameter scaling in the thermodynamic limit.}%

Looking at $\beta_s(s,a_c)$, the parameter $a_c$, or better, the microscopic parameter $g_0$ on which it depends, plays the role that the dimension $d$ plays in a $d$-dimensional Anderson model, see~\cite{altshuler_vanoni_2024}, where it fixes explicitly the position of the critical points and the critical exponent. However, the peculiar form of $\beta_s$ discussed here is such that, despite the fact that the value of $s$ at the critical point changes, the critical exponent $\nu$ stays fixed at 1.

The expression in Eq.~\eqref{def_beta_tps} then allows us to estimate the corrections to the critical exponent as
\begin{eqnarray}\label{eq:nu:corr}
    \nu=\qty(\beta_s'(s_c)s_c)^{-1}=1-\frac{L_0}{L}\;.
\end{eqnarray}
Numerical analysis of the finite-size corrections to $\nu$, shown in Fig.~\ref{figM4} of End Matter, confirm validity of the predicted scaling in Eq.~\eqref{eq:nu:corr} in the regime $g_0>1$.

\textit{Conclusions---}
In this Letter, we studied a new universality class of ergodicity-breaking transitions.
It is described by the framework of fading ergodicity~\cite{kliczkowski_vidmar2024}, i.e., a precursor regime of ergodicity breaking in which volumic fluctuations of observables are gradually transformed to linear fluctuations at the critical point.

The phenomenology of fading ergodicity differs from the one in noninteracting systems that exhibit one-parameter scaling, and possibly also from the well-studied interacting models such as the random-field XXZ model.
We showed that the beta function in the ergodic regime obeys, at large system sizes, the two-parameter scaling.
Our theoretical framework allows for identifying the critical exponent of fading ergodicity $\nu=1$, and it offers a reference point for future explorations of alternative scenarios of ergodicity breaking.

\textit{Acknowledgements---}
We acknowledge discussions with B. Altshuler, A. Chandran, V. Kravtsov and P. Sierant.
R.S., M.H., and L.V. acknowledge support from the Slovenian Research and Innovation Agency (ARIS), Research core funding Grants No. P1-0044, No. N1-0273, No. J1-50005, and No. N1-0369, as well as the Consolidator Grant Boundary-101126364 of the European Research Council (ERC). M. H. acknowledges support from the Polish National Agency for Academic Exchange (NAWA)’s Ulam Programme (project BNI/ULM/2024/1/00124).
C.V. and L.V. acknowledge the Simons Center for Geometry and Physics at Stony Brook for hospitality during part of the work on this project.
The work of A.S. was funded by the European Union - NextGenerationEU under the project NRRP “National Centre for HPC, Big Data and Quantum Computing (HPC)''CN00000013 (CUP D43C22001240001) [MUR Decree n.\ 341- 15/03/2022] - Cascade Call launched by SPOKE 10 POLIMI: “CQEB” project, and by the PNRR MUR project PE0000023-NQSTI..
We gratefully acknowledge the High Performance Computing Research Infrastructure Eastern
Region (HCP RIVR) consortium~\cite{vega1} and European High Performance Computing Joint Undertaking (EuroHPC JU)~\cite{vega2} for funding this research by providing computing resources of the HPC system Vega at the Institute of Information sciences~\cite{vega3}.

\bibliographystyle{biblev1}
\bibliography{references}

\onecolumngrid
\begin{center}
{\large \bf End matter}\\
\end{center}
\twocolumngrid

{\it Appendix A: Quantum sun model---}
The quantum sun model consists of a thermal inclusion (a quantum dot) with $N=3$ spins, and $L$ localized spins outside the dot. The Hamiltonian for this model can be written as
\begin{equation}\label{eq:qsun}
    \hat{H}=\hat{R}+g_0\sum_{\ell=0}^{L-1}\alpha^{u_\ell}\hat{S}^x_{n(\ell)}\hat{S}^x_\ell+\sum_{\ell=0}^{L-1}h_\ell\hat{S}^z_\ell,
\end{equation}
where the thermal dot is modeled by a normalized all-to-all random matrix $\hat{R}$ drawn from the Gaussian orthogonal ensemble (GOE)~\cite{suntajs_deroeck_24, SM}. The spins outside the dot are subject to random fields that act as on-site disorder $h_\ell\in[0.5,1.5]$ drawn from a uniform distribution. 
The interaction term connects a localized spin at site $\ell$ to a randomly selected spin in the dot, $n(\ell)$, via an exponentially decaying strength $g_0\alpha^{u_\ell}$ with $u_\ell\in[\ell-0.2,\ell+0.2]$ and $u_0=1$.
Hence, the distance between the spins inside and outside the dot can be roughly estimated as $u_\ell\sim\ell$.
The last spin, of main interest in this work, is then at distance $u_{L-1}\sim L$.
Decreasing the interaction $\alpha$ drives the model from an ergodic phase at $\alpha>\alpha_c$, through an ergodicity breaking phase transition at $\alpha=\alpha_c$, to a localized phase for $\alpha<\alpha_c$~\cite{suntajs_vidmar_22, suntajs_deroeck_24, hopjan_vidmar_23,hopjan_vidmar_24}.
Following previous theoretical arguments, the transition occurs at $\alpha_c=1/\sqrt{2}\approx0.707$~\cite{deroeck_huveneers_17,luitz_huveneers_17}, while the numerical results suggest the transition to occur at slightly larger $\alpha_c$, which may also depend on $g_0$~\cite{suntajs_deroeck_24}.
We study this model for system sizes up to $L+N=19$, see also~\cite{SM} for further details about the numerical implementation.

\begin{figure}[b!]
\centering
\includegraphics[width=\columnwidth]{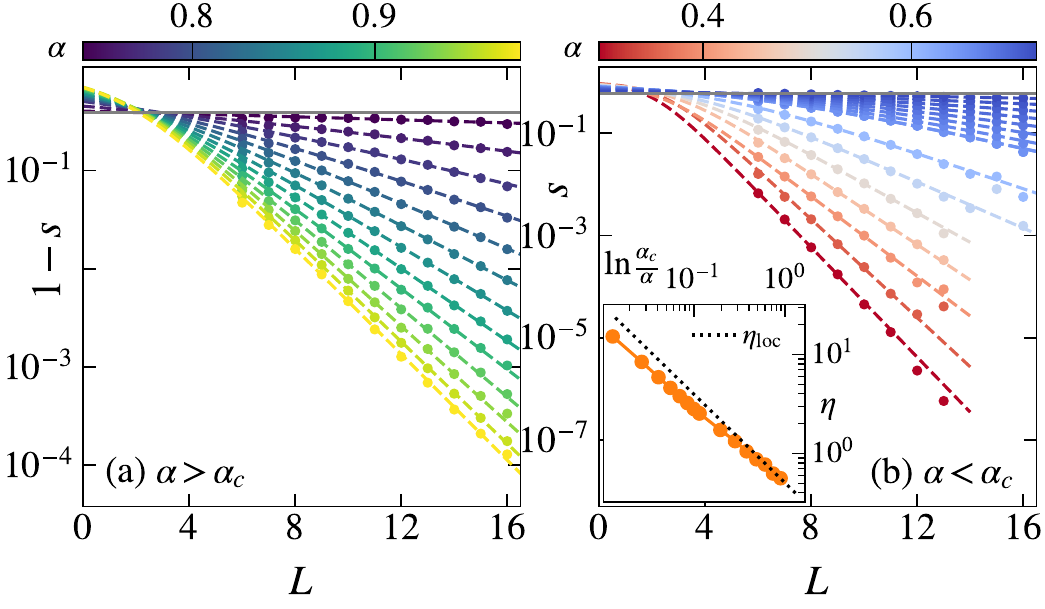}
\caption{
{  (a) Numerical results for $1-s$ vs $L$ in the ergodic phase $\alpha>\alpha_c$,}
and (b) for $s$ vs $L$ in the localized phase $\alpha<\alpha_c$, {  both at $g_0=1$}.
The dashed curves are fits from Eq.~\eqref{eq:s_Fermi} to sizes $L\geq7$. Inset of (b): Correlation length $\eta$ obtained from the fits in the main panel. The dotted line shows the prediction $\eta_{\rm loc}=1/\ln\qty(\alpha_c/\alpha)^2$.
}
\label{figM5}
\end{figure}
\begin{figure}[t!]
\centering
\includegraphics[width=\columnwidth]{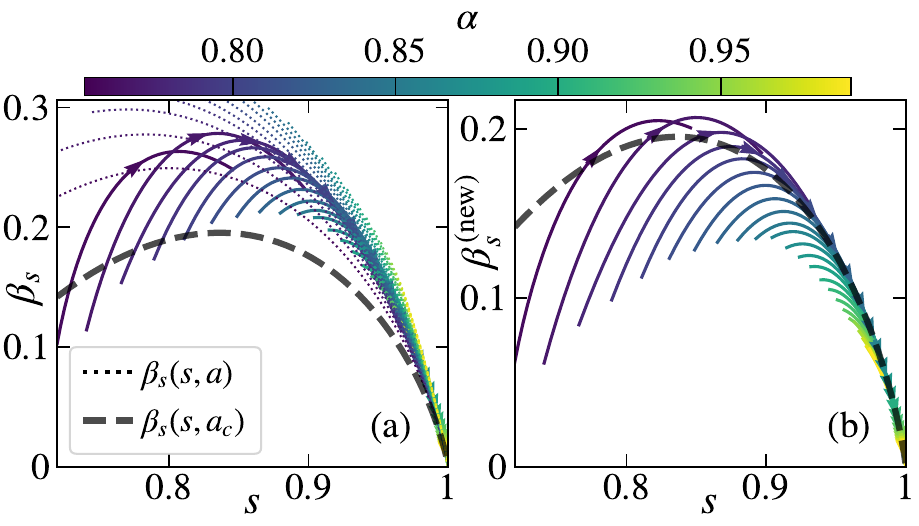}
\caption{  Results for (a) $\beta_s$ and (b) $\beta^{\rm (new)}_s$, in the ergodic phase at $g_0=1$. They complement those in Fig.~\ref{figM3} of the main text.
The thin dotted lines show the two-parameter beta function $\beta_s(s,a)$ from Eq.~\eqref{eq:beta:complete} with values of $a$ extracted from Fig.~\ref{figM5}(a), while the thick dashed line shows the one-parameter function $\beta_s(s,a=a_c)$.
}
\label{figM6}
\end{figure}
\begin{figure}[!t]
\centering
\includegraphics[width=\columnwidth]{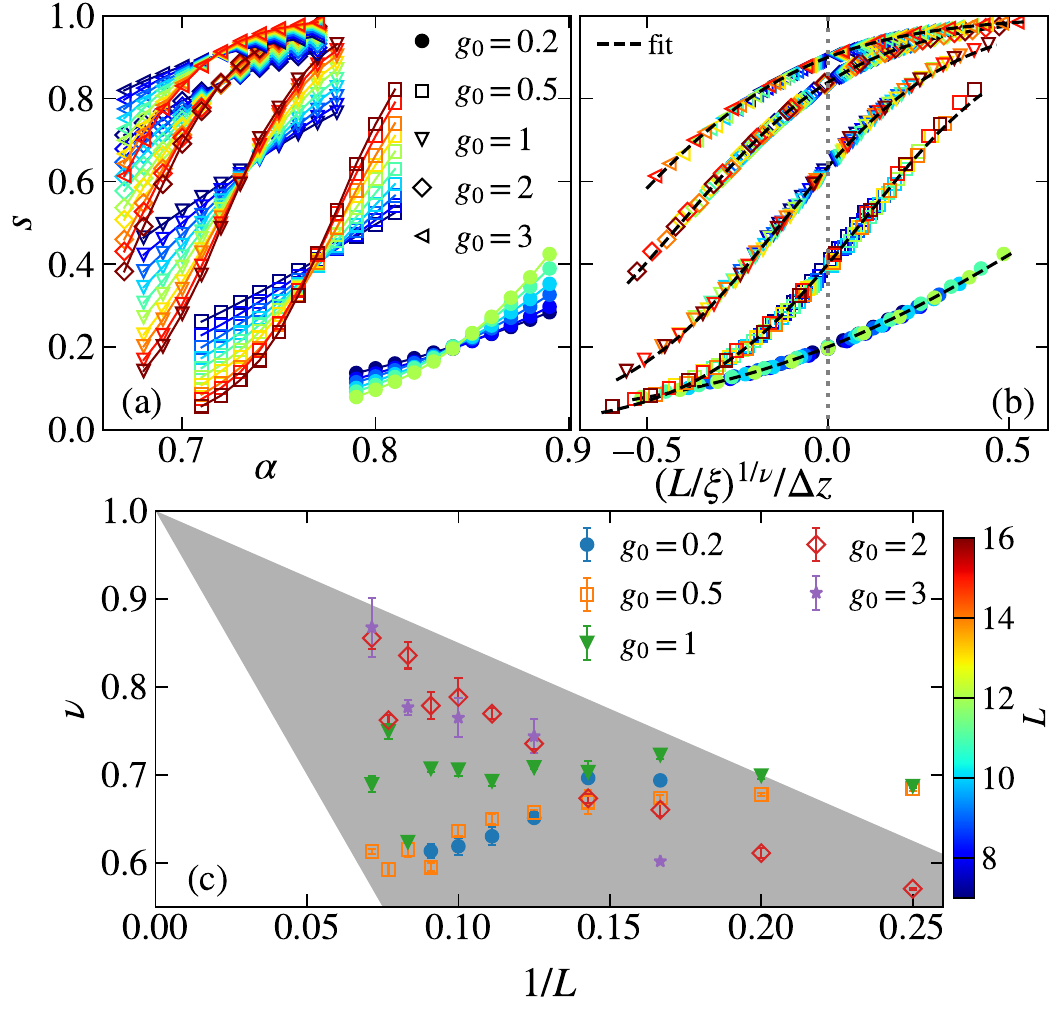}
\caption{
(a) Entanglement entropy $s$ vs $\alpha$ in the vicinity of the critical point, for different values of $L$ and $g_0$.
(b) Scaling collapses using a cost function minimization for the values of $\alpha$ shown in (a).
The dashed line denotes a fit of Eq.~\eqref{eq:s_Fermi} to the collapsed results, \ie upon exchanging $L/\eta\to\qty(L/\xi)^{1/\nu}$. We normalize the values on the $x$ axis by $\Delta z=\max\qty(L/\xi)^{1/\nu}-\min\qty(L/\xi)^{1/\nu}$ to show different $g_0$ on the same scale.
(c) Critical exponent $\nu$ as a function of $1/L$, obtained by the procedure described in the main text.
}
\label{figM4}
\end{figure}

{\it Appendix B: Scaling in ergodic and localized regimes---}
We further support our results in the main text by numerically studying the properties of the beta function in {  the ergodic and} localized phases of the quantum sun model.
{  First, in Fig.~\ref{figM5}(a) we show results for $1-s$ vs $L$ in the ergodic phase at $g_0=1$, which complements the results in Fig.~\ref{figM2} at $g_0=0.5$ and 2, and shows agreement with predictions from the ansatz in Eq.~\eqref{eq:s_Fermi}.} 
In Fig.~\ref{figM5}(b) {  we then show} that the numerical results for $s$ vs $L$ {  in the localized regime} are also in good agreement with the ansatz from Eq.~\eqref{eq:s_Fermi}.
In the inset of Fig.~\ref{figM5}(b), we plot the resulting localization length $\eta$, showing that it agrees with the prediction $\eta_{\rm loc}=1/\ln\qty(\alpha_c/\alpha)^2$ from Ref.~\cite{suntajs_vidmar_22}.
The observed scaling behavior of $s$ is consistent with the beta function in Eq.~\eqref{eq:beta:loc} in the limit $s\to 0$, as argued in the main text.
{  We also show in Fig.~\ref{figM6} the numerical results for the beta function $\beta_s$ from Eq.~\eqref{def_beta_s} and $\beta_s^{\rm (new)}$ from Eq.~\eqref{eq:beta:new}, in the ergodic regime at $g_0=1$.
These results are consistent with those shown and discussed in the context of Fig.~\ref{figM3} in the main text.
}

{\it Appendix C: Scaling collapse at the critical point---}
In Fig.~\ref{figM4}(a) we show additional results for the behavior of $s$ vs $\alpha$ in the vicinity of the critical point for various system sizes $L$ and parameters $g_0$.
As expected, $s_c\to0$ as $g_0\to0$ and $s_c\to1$ as $g_0\to\infty$.
We then find the scaling collapses using the cost function minimization~\cite{suntajs_bonca_20b}, see~\cite{SM} for details.
Here we consider the correlation length $\xi$ consistent with fading ergodicity, \ie $\xi=\eta^\nu\sim[\ln(\alpha/\alpha_c)^2]^{-\nu}$, where $\nu$ is the critical exponent.
While we expect $\nu=1$ in the thermodynamic limit (as argued in the main text), the results in finite systems may give rise to $\nu \neq 1$.

The cost function minimization gives rise to the optimal values of the critical exponent $\nu$, the critical point $\alpha_c$, and the entropy at the critical point $s_c$.
The resulting scaling collapses are shown in Fig.~\ref{figM4}(b), while the values of $\nu$, $\alpha_c$ and $s_c$, as well as the quality of the scaling collapses, are listed in Sec.~S3 of~\cite{SM}.
In general, we find better scaling collapses at large $g_0$, for which the critical point $\alpha_c$ is closer to the analytically predicted value $\alpha_c=1/\sqrt{2}$, and $\nu$ is closer to $\nu=1$.
In the numerical analysis, $\nu$ is smaller than the theoretically predicted $\nu=1$.
Using~\eqref{eq:nu:corr}, we estimate $\nu=1-L_0/L\in(0.6,0.8)$ [since $L_0\in(3,6)$ in Fig.~\ref{figM2}], which is in good agreement with the values given from the scaling collapses.

We complement our analysis by estimating the corrections to the critical exponent $\nu$, namely,
we calculate the critical exponent as a function of system size. 
For a given $L$, we perform a finite-size collapse using system sizes in the range $\qty[L-1,L+2]$ and vary $L$ across the available data points. We use the same correlation length $\xi=\eta^\nu$ as in Fig.~\ref{figM4}(b). The resulting critical exponents are shown in Fig.~\ref{figM4}(c) as a function of $1/L$. For large values of $g_0>1$ we find remarkable agreement with our framework described in the main text.
We estimate $L_0\sim3$ from the results.
However, smaller values of $g_0<1$ show larger finite-size effects and it is not clear whether the flow of the critical exponent $\nu$ will resume $1-L_0/L$ corrections at large $L$, or whether $\nu$ remains smaller, giving rise to $\nu<1$ in the thermodynamic limit.

\clearpage

\newpage
\phantom{a}
\newpage
\setcounter{figure}{0}
\setcounter{equation}{0}
\setcounter{table}{0}
\setcounter{section}{0}

\renewcommand{\thetable}{S\arabic{table}}
\renewcommand{\thefigure}{S\arabic{figure}}
\renewcommand{\theequation}{S\arabic{equation}}
\renewcommand{\thepage}{S\arabic{page}}

\renewcommand{\thesection}{S\arabic{section}}

\onecolumngrid

\begin{center}

{\large \bf Supplemental Material: Scaling theory of fading ergodicity}\\

\begin{center}
Rafał Świętek$^{1,2}$, Miroslav Hopjan$^{1,3}$, Carlo Vanoni$^{4,5}$, Antonello Scardicchio$^{6,7}$ and Lev Vidmar$^{1,2}$\\
$^1${\it Department of Theoretical Physics, J. Stefan Institute, SI-1000 Ljubljana, Slovenia} \\
$^2${\it Department of Physics, Faculty of Mathematics and Physics, University of Ljubljana, SI-1000 Ljubljana, Slovenia} \\
$^3${\it Institute of Theoretical Physics, Wroclaw University of Science and Technology, 50-370 Wrocław, Poland} \\
$^4${\it SISSA – International School for Advanced Studies, via Bonomea 265, 34136, Trieste, Italy}\\
$^5${\it Department of Physics, Princeton University, Princeton, New Jersey, 08544, USA}\\
$^6${\it INFN Sezione di Trieste, Via Valerio 2, 34127 Trieste, Italy}\\
$^7${\it ICTP, Strada Costiera 11, 34151, Trieste, Italy}\\

\end{center}

\vspace{0.3cm}

\setcounter{page}{1}

\end{center}

\vspace{0.6cm}

\twocolumngrid

\label{pagesupp}

\section{Numerical implementation of the model}\label{app:technical}
Here we comment on the numerical implementation of the quantum sun model Hamiltonian defined in Eq.~\eqref{eq:qsun} of End Matter. 
We model the thermal quantum dot with a normalized $2^N\times 2^N$ matrix from a Gaussian orthogonal ensemble (GOE), defined as
\begin{equation}\label{eq:sm:goe:normalized}
    R=\frac{R_0}{\sqrt{2^N+1}}\;,
\end{equation}
where $R_0=(A+A^T)/2$, with $A$ being a random matrix with elements distributed from a normal distribution. 
The matrix $R$ from Eq.~\eqref{eq:sm:goe:normalized} has a unit Hilbert-Schmidt norm~\cite{suntajs_deroeck_24}.
Then, the corresponding operator of the Hamiltonian is a tensor product of $R$ with an identity operator acting on $L$ spins outside the dot, \ie $\hat{R}=R\otimes\mathcal{I}$.
The spins outside the dot are coupled to a randomly selected spin within the dot.
We set the coupling of the first spin outside the dot to $\alpha^{u_0}=1$, \ie $u_0=0$, while the remaining $u_{\ell\geq1}$ are drawn uniformly from the interval $[\ell-0.2,\ell+0.2]$.

Numerical results for the entanglement entropy $s$ are obtained by carrying out an average over $\mathcal{N}_{av}$ Hamiltonian realizations.
We use $\mathcal{N}_{av}\geq5000$ realizations for $L+N\leq12$, $\mathcal{N}_{av}=4000$ realizations for $L+N=13,14$, $\mathcal{N}_{av}\geq1500$ disorder realizations for $L+N=15$, and $\mathcal{N}_{av}\geq500$ for $L+N=16$.
For system sizes beyond full exact diagonalization (ED) we employ polynomially filtered ED (POLFED)~\cite{sierant_lewenstein_20} to target $N_{eig}=500$ central eigenvalues with a tolerance of $\delta=10^{-14}$.
In this case we use $\mathcal{N}_{av}\geq500$ realizations for $L+N=17$ and 18, while for $L+N=19$ we perform $\mathcal{N}_{av}\geq200$ averages.

\section{Entanglement entropy vs matrix elements} \label{sec:sm:matele}

Let us write a general state $\ket{\psi}$ of a system of $L$ spins-1/2 in the basis of product states that are eigenstates of the local magnetization operator $\hat{S}^z_\ell$,
\begin{equation}
    \ket{\psi}=\sum_{k=1}^Dc_k\ket{k},
\end{equation}
where $\ket{k}\in(\mathbb{C}^2)^{\otimes^L}$ are the basis states and $c_k=\braket{k}{\psi}$ is the coefficient of the wave function. In this work, we are interested in the properties of a single spin. The basis $\qty{\ket{k}}$ can be decomposed in the basis $\ket{b}\otimes\ket{\sigma}$, where $\ket{b}\in(\mathbb{C}^2)^{\otimes^{L-1}}$ and $\ket{\sigma}\in\mathbb{C}^2=\qty{\ket{\downarrow},\ket{\uparrow}}$. Hence, our state takes the form
\begin{equation}
    \ket{\psi}=\sum_{b}\sum_{\sigma}c_{b\sigma}\ket{b}\otimes\ket{\sigma}.
\end{equation}
Next, we can find the reduced density matrix for the single spin on site $\ell$ and trace out the remaining degrees of freedom, \ie
\begin{equation}
    \begin{split}
        \hat{\rho}_\ell&=\Tr_B\dyad{\psi}\\
        &=\sum_{b"}\sum_{b,b'}\sum_{\sigma,\sigma'}c_{b\sigma}c^*_{b'\sigma'}\mel{b"}{\qty(\ket{b}\otimes\ket{\sigma}\bra{\sigma'}\otimes\bra{b'})}{b"}\\
        &=\sum_{\sigma,\sigma'}\sum_bc_{b\sigma}c^*_{b\sigma'}\ketbra{\sigma}{\sigma'}=\mqty(A_{00} & A_{01}\\ A_{10} & A_{11}),
    \end{split}
\end{equation}
where $A_{\sigma\sigma'}=\sum_bc_{b\sigma}c^*_{b\sigma'}$. Due to the property $\Tr{\hat\rho_\ell}=1$, we have $A_{11}=1-A_{00}$ and we set $A_{01}=A_{10}^*$ due to hermiticity of the density matrix.
The eigenvalues of the matrix $\hat{\rho}_\ell$ can be then expressed as
\begin{equation}
    \lambda_{\pm}=\frac{1}{2}\pm\frac{1}{2}\sqrt{\qty(1-2A_{00})^2+4\abs{A_{01}}^2)}.
\end{equation}
On the other hand, the expectation value $z$ of the local magnetization $\hat{S}^z_\ell$ can be found in a similar manner,
\begin{equation}
   \begin{split}
        z \equiv\mel{\psi}{\hat{S}^z_\ell}{\psi}&=\sum_{b,b'}\sum_{\sigma,\sigma'}c^*_{b'\sigma'}c_{b\sigma}\bra{b'}\otimes\bra{\sigma'}\hat{S}^z_\ell\ket{\sigma}\otimes\ket{b}\\
        &=\sum_{\sigma,\sigma'}\sum_bc^*_{b\sigma'}c_{b\sigma}\underbrace{\mel{\sigma'}{\hat{S}^z_\ell}{\sigma}}_{\sigma\cdot\delta{\sigma\sigma'}}\\
        &=\frac{1}{2}A_{11}-\frac{1}{2}A_{00}=\frac{1}{2}-A_{00}\;,
   \end{split}
\end{equation}
where we choose the spin-$\frac{1}{2}$ eigenvalues $\sigma\in\qty{-\frac{1}{2},\frac{1}{2}}$. Following a similar calculation one can additionally find that the off-diagonal elements of the reduced density matrix become
\begin{equation}
    A_{01}=\mel{\psi}{\hat{S}^-_\ell}{\psi}=\mel{\psi}{\hat{S}^x_\ell-i\hat{S}^y_\ell}{\psi} \equiv x -iy.
\end{equation}
Remarkably, this indicates that the eigenvalues $\lambda_{\pm}$ of the reduced density matrix are related to the fluctuations of matrix elements at the same site,
\begin{equation}
    \lambda_{\pm}=\frac{1}{2}\qty(1\pm2\sqrt{\abs{x}^2+\abs{y}^2+\abs{z}^2}).
\end{equation}

For a $U(1)$ symmetric system, the off-diagonals of the reduced density matrix disappear, \ie $x=y=0$. We can then write the single-site entanglement entropy $S$ for the state $\ket{\psi}$ as
\begin{equation}
    \begin{split}
        S &=-\Tr\qty(\hat\rho_\ell\ln\hat\rho_\ell)\\
        &=-\qty(\frac{1}{2}-\frac{z}{4})\ln{\qty(\frac{1}{2}-\frac{z}{4})}-\qty(\frac{1}{2}+\frac{z}{4})\ln{\qty(\frac{1}{2}+\frac{z}{4})}\;.
    \end{split}
\end{equation}
In the ETH regime at infinite temperature, one can assume a vanishingly small value of the magnetization, such that we can expand the above expression at $z=0$ to get
\begin{equation} \label{def_S_z_correction}
    S\approx\ln{2}-2z^2+O(z^4).
\end{equation}
Consequently, we find that the average of entanglement entropies over some states $\psi$ (e.g., the Hamiltonian eigenstates) are directly related to the variance of diagonal matrix elements over the same set of states. 

Similar considerations apply for models without $U(1)$ symmetry, where the off-diagonal terms of the reduced density matrix $A_{01}$ modify Eq.~\eqref{def_S_z_correction} to
\begin{equation} \label{def_S_u_z}
    \begin{split}
        S\approx\,&\abs{u}\ln\frac{1-2\abs{u}}{1+2\abs{u}}-\frac{1}{2}\ln{\qty(\frac{1}{4}-\abs{u}^2)}\\
        &+\frac{1}{2\abs{u}}\ln{\qty(\frac{1-2\abs{u}}{1+2\abs{u}})}z^2+O(z^4),
    \end{split}
\end{equation}
with $u=A_{01}$.
As argued in~\cite{kliczkowski_vidmar2024}, fading ergodicity manifests itself for certain physical observables, such as $\hat S_\ell^z$, as softening of fluctuations of matrix elements, i.e., the matrix elements fluctuations are larger than those predicted by the ETH.
On the other hand, other observables such as those that include the operator $\hat S_\ell^x$ still exhibit faster (ETH-like) decay of fluctuations.
Hence, we expect fluctuations of $S$ in Eq.~\eqref{def_S_u_z} to be governed by $z$.
Sending $u\to 0$ in Eq.~\eqref{def_S_u_z} while keeping $z>0$ gives rise to the same expression as given in Eq.~\eqref{def_S_z_correction}.

\section{Finite-size scaling of the entanglement entropy}

\subsection{Results for random pure states} \label{app:Wishart}

Within the random matrix theory, the distribution for the eigenvalues of the reduced density matrix is the Wishart distribution, which for a generic parameter $\beta$ (not to be confused with the beta function) takes the form~\cite{James1964Distribution} 
\begin{equation}
    \begin{split}
    d\mu =\; & \frac{1}{Z_{\beta}} d^N \lambda \prod_{i<j}|\lambda_i - \lambda_j|^{\beta} \\
    &\times \; \prod_l \lambda_l^{[(1+M-N)-2/\beta]\beta/2}\; \delta(1-\sum_k \lambda_k)\;,
    \end{split}
\end{equation}
where the SO(N) volume has been dropped and $\lambda_i>0$,  $\forall \, i$.
In our case, we have to consider a GOE matrix ($\beta = 1$) of dimension $\mathcal{D}$, where the reduced density matrices of the subsystems have dimension $N=2$ and $M=\mathcal{D}/N$.
This gives
\begin{equation}\label{eq:sm:entropy:measure}
    d\mu = \frac{1}{Z} d \lambda_1 d \lambda_2 |\lambda_1 - \lambda_2|  \lambda_1^{\frac{\mathcal{D}-6}{4}} \lambda_2^{\frac{\mathcal{D}-6}{4}}\delta(1- \lambda_1 - \lambda_2)\;,
\end{equation}
with
\begin{equation}
    Z = \frac{2^{6-\mathcal{D}}}{\mathcal{D}-2}\;.
\end{equation}
The entanglement entropy is then $s=-(\lambda_1\ln\lambda_1+\lambda_2\ln\lambda_2)/\ln{2}$, so the average can be computed as
\begin{equation}
    \expval{s}=-\frac{2}{\ln{2}}\int d\mu \lambda_1\ln(\lambda_1).
\end{equation}

Since the measure $d\mu$ in Eq.~\eqref{eq:sm:entropy:measure} contains an absolute value, it is convenient to split the integral into two parts, $\expval{s} = 2(I_1+I_2)/\ln{2}$, of the form
\begin{equation}
    I_1 = -\frac{1}{Z} \int_{0}^{1/2}d\lambda (1-2\lambda)(\lambda-\lambda^2)^n \lambda\ln \lambda
\end{equation}
and
\begin{equation}
    I_2 = -\frac{1}{Z} \int_{1/2}^{1}d\lambda (2\lambda-1)(\lambda-\lambda^2)^n \lambda\ln \lambda\;,
\end{equation}
where we also set $n = (\mathcal{D}-6)/4$. To solve the integrals we introduce a change of variables, such that we can recast $I_1$ as
\begin{equation}
    I_1 = -\frac{n+1}{2}\int_{0}^{1} dy \, y^n \lambda_-(y) \ln \lambda_-(y),
\end{equation}
and $I_2$ as
\begin{equation}
    I_2 = -\frac{n+1}{2}\int_{0}^{1} dy \, y^n \lambda_+(y) \ln \lambda_+(y),
\end{equation}
where $\lambda_{\pm} = (1\pm \sqrt{1-y})/2$. In the large $n$ limit (corresponding to the large $\mathcal{D}$ limit), both integrals are dominated by the contribution near $y=1$, so we can expand in the small parameter $\epsilon = 1-y$. In the expansion, the odd powers in $I_1$ and $I_2$ cancel, while the even powers sum and one gets, for $\mathcal{D}\gg1$,
\begin{align}\label{eq:sm:entropy:rmt}
    \expval{s} = 1 -\frac{\frac{2}{\ln{2}}}{\mathcal{D}} + O\left( \frac{1}{\mathcal{D}^2} \right).
\end{align}
When considering a physical Hamiltonian instead of the GOE matrix, ${\cal D} \to D$ is the Hilbert-space dimension.
Then, the leading term of the random matrix theory prediction for the fluctuations of the entanglement entropy is proportional to $1/D$, as argued in the main text.

\begin{figure}[t]
\centering
\includegraphics[width=\columnwidth]{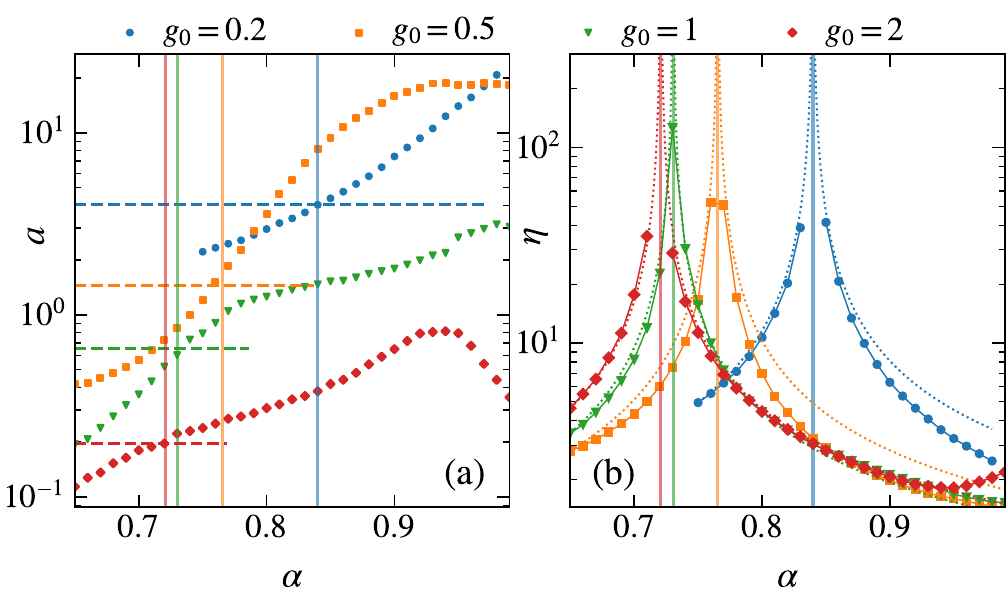}
\caption{(a) and (b)
Fitting constants $a$ and $\eta$ from Eq.~\eqref{eq:sm:sL}, respectively, for different values of $g_0$ as a function of $\alpha$.
The vertical solid lines show the critical interaction $\alpha_c$, while the horizontal dashed lines in (a) show $a_c=1/s_c-1$ for different values of $g_0$ (marked by the same color), extracted from the data collapse in the main text (see End Matter).
The dotted lines in (b) show the fit of $\bar\eta(\alpha)=b\abs{\ln\qty(\alpha/\alpha_c)}^{-1}$ to the numerical results for $\eta$ vs $\alpha$, where $b$ is a fitting parameter.
}
\label{figS1}
\end{figure}

\subsection{Numerical tests for the scaling ansatz}
 \label{sec:numerical_test}
Here we test the ansatz for the finite-size scaling of the entanglement entropy in the quantum sun model,
\begin{eqnarray} \label{eq:sm:sL}
s(L)= \frac{1}{1+a e^{-L/\eta}} \;,
\end{eqnarray}
which is Eq.~\eqref{eq:s_Fermi} in the main text. 
We show the results, at different values of $g_0$, for $a$ vs $\alpha$ in Fig.~\ref{figS1}(a) and $\eta$ vs $\alpha$ in Fig.~\ref{figS1}(b).

In Fig.~\ref{figS1}(a) we find $a\to a_c=1/s_c-1$ in the vicinity of the critical point, while $a>a_c$ in the ergodic regime.
In Fig.~\ref{figS1}(b) we find a clear tendency of the divergence of $\eta$ in the vicinity of the critical point.
We fit the numerical results for $\eta$ with an ansatz obtained from the analysis of the matrix elements fluctuations of the operator $\hat S^z_L$ in the fading ergodicity regime, see Fig.~3(d) of~\cite{kliczkowski_vidmar2024}.
The agreement between the numerical results and the fitted function is remarkable, suggesting the similarity between the scaling properties of the variances of the matrix elements of $\hat S^z_L$ and the eigenstate entanglement entropy $s$, as anticipated from the analysis in Sec.~\ref{sec:sm:matele}.

\subsection{Scaling collapses in the critical regime}\label{app:collapse}

In Fig.~\ref{figM4}(a) of the main text (see End Matter) we show results for $s$ vs $\alpha$ in the vicinity of the critical point, at various $L$ and $g_0$.
The maximal system size under investigation depends on the value of $g_0$:
for $g_0<0.5$ we study up to $L+N=15$ particles, for $g_0=3$ we study up to $L+N=18$ particles, while for $g_0=0.5,1,2$ we study up to $L+N=19$ particles.

\begin{table}[!t]
\centering
\begin{tabular}{|p{1.25cm}| |p{1.25cm}||p{1.25cm}|p{1.25cm}|p{1.25cm}|}
    \hline
	$g_0$& $C_X$ & $\alpha_c$ & $s_c$ & $\nu$\\[8pt]
	\hline\hline
	$0.5$ & $0.354$ & $0.768$ & $0.408$ & $0.597$\\[8pt]
 \hline
	$1$ & $0.146$ & $0.734$ & $0.605$ & $0.691$\\[8pt]
 \hline
	$2$ & $0.065$ & $0.721$ & $0.836$ & $0.790$\\[8pt]
 \hline
	$3$ & $0.033$ & $0.716$ & $0.909$ & $0.807$\\[8pt]
 \hline
\end{tabular}
\caption{Summary of cost function analysis, see Eq.~(\ref{eq:sm:costfun}), for $s$ vs $\alpha$ and different $L$.
The scaling collapses are shown in Fig.~\ref{figM4}(b) of the main text.
For a fixed $g_0$, the fitting parameters are $\alpha_c$, $s_c$ and $\nu$.
\label{tab1}}
\end{table}

In Fig.~\ref{figM4}(b) we show the scaling collapses of the results in Fig.~\ref{figM4}(a) obtained by using the cost function minimization from Ref.~\cite{suntajs_bonca_20b}. 
Here we briefly summarize the main steps of the cost function minimization.
The cost function for the observable $X$ is defined as
\begin{equation}\label{eq:sm:costfun}
    C_X=\frac{\sum_i\abs{X_{i+1}-X_i}}{\max X - \min X}-1,
\end{equation}
such that the optimal collapse is given by $C_X=0$.
To carry out the cost function procedure, we employ the {\tt differential evolution} algorithm from the Python {\tt scipy.optimize} library. We utilize a population size of $10^2$ with up to $10^3$ iterations, and an absolute tolerance of $10^{-2}$. Given the stochastic nature of this method, we perform over $100$ realizations and use the typical values (i.e., geometric mean) of the fitting parameters. 

The results of the cost function minimization are summarized in Table~\ref{tab1} for $g_0=0.5,1,2,3$, which are the parameters for which we considered the largest system sizes.
The best scaling collapses are found for large $g_0$, for which the critical point $\alpha_c$ is closest to the analytically predicted value $\alpha_c=1/\sqrt{2}$, and the critical exponent $\nu$ is closest to $\nu=1$.
These results reinforce our expectation in the discussion of Fig.~\ref{figM3} in the main text, namely, that the numerical results at large $g_0$ are closest to the asymptotic regime, while the results at small $g_0$ are subject to strong finite-size effects.

\begin{figure}[t!]
\centering
\includegraphics[width=\columnwidth]{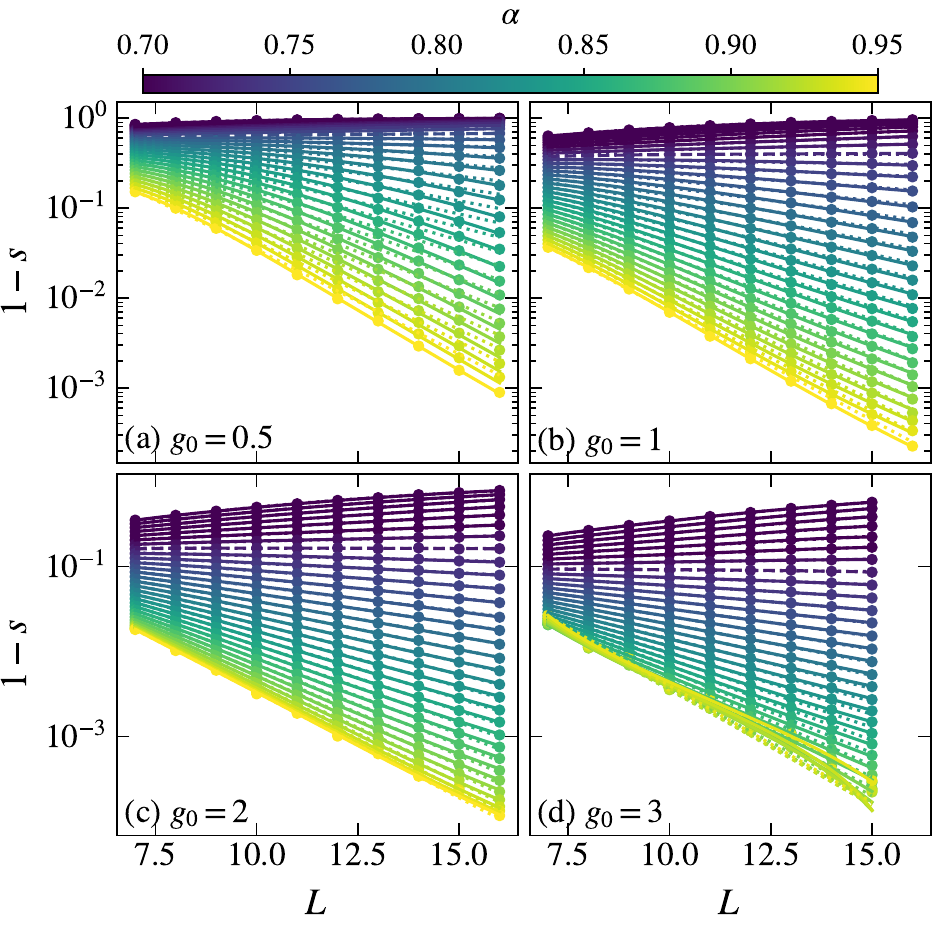}
\caption{
Finite-size scaling of the subtracted entanglement entropy $1-s$ vs $L$, for different values of $g_0$. The solid lines represent the fits of Eqs.~\eqref{eq:sm:fit:erg} and~\eqref{eq:sm:fit:loc} using $x=\exp{L/\eta}$, while the dotted lines show the fit of Eq.~\eqref{eq:sm:sL} to the results.
The horizontal dashed line is the estimated critical point.
}
\label{figS2}
\end{figure}

\section{Numerical calculation of the beta function}
In our study in the main text, we consider finite systems of sizes up to $L+N=19$.
To ensure a smooth curve for the entanglement entropy, $s(L)$, we fit our results in the ergodic regime (at $\alpha>\alpha_c$) to the Pade function $F_1$,
\begin{equation}\label{eq:sm:fit:erg}
    F_1(x)=\frac{x^2+c_1x+c_0}{x^2+d_1x+d_0},
\end{equation}
which ensures that $s\xrightarrow[]{L\rightarrow\infty}1$, and in the localized regime (at $\alpha<\alpha_c$) to the Pade function $F_2$,
\begin{equation}\label{eq:sm:fit:loc}
    F_2(x)=\frac{c_3}{x^3}+\frac{c_2}{x^2}+\frac{c_1}{x}+\frac{d}{\sqrt{x}} + c_0,
\end{equation}
such that $s\xrightarrow[]{L\rightarrow\infty}{\rm const}$.
Since the corrections in the ergodic and localized regimes are exponential, the fitting variable should be $x\sim\exp{L}$. However, taking into account the fading ergodicity regime, a more relevant fitting parameter should be $x=\exp{L/\eta}$.

Following this method we are able to obtain smooth logarithmic derivatives of the interpolated results, which is central to the numerical calculation of the beta function.
We fit the functions in Eqs.~\eqref{eq:sm:fit:erg} and~\eqref{eq:sm:fit:loc} to our results for $s$ with $x=\exp{L/\eta}$, and we show the results in Fig.~\ref{figS2}(a)-\ref{figS2}(d) for the whole range of $\alpha$ and different $g_0$.
Remarkably, the interpolated results fit the data points very well and reduce fluctuations due to the finite number of realizations. Alternatively, we fit Eq.~\eqref{eq:sm:sL} to our results for all system sizes to test the validity of the ansatz. While in the asymptotic limit the fit seems to work remarkably well (see Fig.~\ref{figM2} of the main text), here we note that using Eq.~\eqref{eq:sm:fit:erg} we obtain a more robust fitting function for moderate system sizes. A natural explanation is that the expansion of Eq.~\eqref{eq:sm:fit:erg} contains corrections beyond Eq.~\eqref{eq:sm:sL}, which become relevant for smaller system sizes. Therefore, to evaluate the beta function it is more robust to choose the Pade function instead of the asymptotic scaling function expressed via Eq.~\eqref{eq:sm:sL}.


\end{document}